\documentclass[10pt]{article}

\usepackage[dviwin]{graphicx}
\usepackage{amsfonts}
\usepackage{amssymb}
\usepackage{mathrsfs}
\usepackage{amsmath}
\usepackage{indentfirst}
\usepackage{mathrsfs}

\usepackage{dsfont}
\usepackage{esint}
\usepackage{hyperref}
\hypersetup{
    colorlinks = true,
    linkcolor = [rgb]{0 .4 .8},
    citecolor=[rgb]{1 .2 .2}
}
\usepackage{array}

\usepackage[numbers,sort]{natbib}
\setcitestyle{square}

\newcommand{\bea}{\begin{equation}}
\newcommand{\eea}{\end{equation}}
\newcommand{\bear}{\begin{eqnarray}}
\newcommand{\eear}{\end{eqnarray}}
\newcommand{\bearr}{\begin{eqnarray*}}
\newcommand{\eearr}{\end{eqnarray*}}

\newcommand{\beal}{\begin{align}}
\newcommand{\eeal}{\end{align}}
\newcommand{\beall}{\begin{align*}}
\newcommand{\eeall}{\end{align*}}

\newcommand{\tr}{\mathrm{tr}\,}
\newcommand{\CP}{\mathds{C}\mathds{P}}
\newcommand{\CC}{\mathds{C}}

\newcommand{\dd}{\partial}

\newcommand{\comment}[1]{}
\newcommand{\gfr}{\mathfrak{g}}
\newcommand{\mg}{\mathfrak{m}}

\usepackage{xcolor}
\usepackage{empheq}

\newcolumntype{L}[1]{>{\raggedright\let\newline\\\arraybackslash\hspace{0pt}}m{#1}}
\newcolumntype{C}[1]{>{\centering\let\newline\\\arraybackslash\hspace{0pt}}m{#1}}
\newcolumntype{R}[1]{>{\raggedleft\let\newline\\\arraybackslash\hspace{0pt}}m{#1}}

\makeatletter
\def\@seccntformat#1{\@ifundefined{#1@cntformat}%
{\csname the#1\endcsname\quad}% default
{\csname #1@cntformat\endcsname}% individual control
}
\def\section@cntformat{{\normalfont\Large\thesection.}\quad}
\def\subsection@cntformat{\textsection\, \thesubsection.\quad}
\def\subsubsection@cntformat{\textsection\textsection\, \thesubsubsection.\quad}
\makeatother

\newsavebox\MBox

\newcommand*{\TitleFont}{%
      \usefont{\encodingdefault}{\rmdefault}{b}{n}%
      \fontsize{14}{20}%
      \selectfont}
 
\usepackage{sectsty}
 
\usepackage{bm}

\DeclareBoldMathCommand\boldlangle{\left\langle}
\DeclareBoldMathCommand\boldrangle{\right\rangle}

\sectionfont{\fontsize{12}{15}\selectfont}

\begin{document}
\vspace{-1cm}
\title{\vspace{-1.5cm}\TitleFont Complex structure-induced deformations of $\sigma$-models}
\author{Dmitri Bykov\footnote{Emails:
dmitri.bykov@aei.mpg.de, dbykov@mi.ras.ru}  \\ \\
{\small $\bullet$ Max-Planck-Institut f\"ur Gravitationsphysik, Albert-Einstein-Institut,} \\ {\small Am M\"uhlenberg 1, D-14476 Potsdam-Golm, Germany} \\ {\small $\bullet$ Steklov
Mathematical Institute of Russ. Acad. Sci.,}\\ {\small Gubkina str. 8, 119991 Moscow, Russia \;}}
%\date{18.05.2013 / ver. 27.06.2013}
\date{}

\maketitle
\vspace{-1cm}
\begin{center}
\line(1,0){370}
\end{center}
\vspace{-0.2cm}
\textbf{Abstract.} We describe a deformation of the principal chiral model (with an even-dimensional target space $G$) by a $B$-field proportional to the K\"ahler form on the target space. The equations of motion of the deformed model admit a zero-curvature representation. As a simplest example, we consider the case of $G=S^1\times S^3$. We also apply a variant of the construction to a deformation of the $AdS_3\times S^3 \times S^1$ (super-)$\sigma$-model.

\vspace{-0.7cm}
\begin{center}
\line(1,0){370}
\end{center}

The main goal of this paper is to relate the $\sigma$-models introduced by the author \cite{Bykov1}, \cite{Bykov2}, \cite{Bykov3} to the so-called $\eta$-deformed models \cite{Klimcik}, which have recently attracted considerable interest \cite{Vicedo}, \cite{ABF}. The relation is based on the interpretation of the $R$-matrix of the latter models as a complex structure on the target space of the $\sigma$-model. In this case the $R$-matrix has eigenvalues $\pm i$ and is therefore non-degenerate. This is in contrast to the $R$-matrix utilized in \cite{Klimcik}, \cite{Vicedo}, where it has a nontrivial null space.

The relation that we find between the two classes of models is \emph{not} one-to-one. Let us explain this. First of all, the models of \cite{Bykov3} do not have any free parameters -- in this sense they are not deformations of any simpler models. However, in certain cases they may be obtained as limits of the $\eta$-deformed models for special values of $\eta$ (In standard normalization, this is $\eta = \pm i$). For instance, the $\eta$-deformed model with target space $SU(N)$ degenerates in this limit to a $\sigma$-model with target space $SU(N)/S(U(1)^N)$ -- the \emph{complete} flag manifold. Two remarks are in order:
\begin{itemize}
\item The target spaces of $\sigma$-models obtained in such limit are always of the type $G/H$ with \emph{abelian} $H$. On the other hand, the models of \cite{Bykov3} are defined for arbitrary complex homogeneous spaces, irrespective of whether $H$ is abelian. They are only well-defined, however, for Euclidean worldsheets (From the point of view of the limit, the reason is that $\eta$ needs to be taken complex).
\item The inverse procedure does not exist in general, i.e. there is in general no $\eta$-deformation of the flag manifold $\sigma$-model. The reason is that the limit $\eta \to \pm i$ in general irreversibly modifies the target space of the model. When it does not, the $R$-matrix deformation provides generalizations of the models of \cite{Bykov3}. This is so, for instance, when the target space is a group manifold, and the $R$-matrix is a complex structure on this group -- then the $\eta$-deformed model interpolates between the model of \cite{Bykov3} and the principal chiral model. In this case the deformation amounts to adding a $B$-field proportional to the (non-closed) K\"ahler form on the target space.
\end{itemize}
Despite the fact that the technical side of the theory of $R$-deformation is rather involved, the deformation itself is of conceptual interest. One interesting feature is the relation of the deformation to the existence of two Poisson brackets on the space of fields of a $\sigma$-model: the one arising in a canonical way from the original theory and the other one arising from its Pohlmeyer-reduced version. This line of thought was pursued in \cite{VicedoSymm}, but it will not be discussed in the present paper. Below the emphasis will be put instead on the relation of the deformation to the complex structures on the target space.

\vspace{0.3cm}
The structure of the paper is as follows:

Section \ref{general}. We review the one-parametric deformation of the principal chiral model (the one preserving left invariance of the action); an analogous deformation of symmetric space $\sigma$-models in Section \ref{symmspacesec}; two-parametric deformations of the principal chiral model in Section \ref{2param}. We make an observation that, in the case of a one-parametric deformation of the principal chiral model, the Noether current of the model (associated to the remaining left symmetries) is flat.

Section \ref{Rmatrix}. We discuss two possible choices of the $R$-matrix satisfying the `modified classical Yang-Baxter equation'. We start from the possibility elaborated most in the literature, namely an $R$-matrix that is allowed to have zero eigenvalues. In Section~\ref{limitcase} we show that in the limit $\eta \to \pm i$ one recovers some models of  the class considered in~\cite{Bykov3}. Another possibility corresponds to choosing an integrable complex structure $\mathscr{J}$ as $R$-matrix. In Section \ref{complstruct} we recall that there exists an integrable complex structure on a compact even-dimensional Lie group.

Section \ref{S1S3}. This section is dedicated to the analysis of the simplest complex-structure deformed principal chiral model -- the $\sigma$-model with target space $S^1\times S^3$. In Sections \ref{R4complstruct}, \ref{hyperhermit} we review the complex structures on $\mathbb{R}^4$ and on $S^1\times S^3$. We then write out the Lagrangian of the model in Section \ref{Themodel} and discuss the $T$-dual target space, as well as the mechanical reduction of the model.

Section \ref{AdSDef}. We apply the complex structure-induced deformation of the $S^1\times S^3$ model to an analogous deformation of the $AdS_3\times S^3 \times (S^1)^4$ $\sigma$-model, which is of interest in the framework of the AdS/CFT correspondence. This model additionally contains fermions, which make it supersymmetric in the target space. Since the symmetry group $PSU(1,1|2)\times PSU(1,1|2)\times U(1)^4$ of the undeformed theory mixes the $AdS_3$ and $S^3$ parts, the corresponding deformation of the $S^3$ part necessarily induces a deformation of the $AdS_3$ part as well. We show that this `deformation' of $AdS_3$ is in fact a completely different background, which in particular has closed timelike curves.

\section{\Large The $\eta$-deformation}\label{general}

In \cite{Klimcik} a certain deformation of the principal chiral model with target space $G$ was proposed. Let us start by considering the simplest, one-parametric, version of this deformation. It is given in terms of a linear operator $R$ on the Lie algebra $\mathfrak{g}$:
\bea\label{Asymm}
R:\quad \mathfrak{g} \to \mathfrak{g}
\eea
One requires that the Lie algebra $\mathfrak{g}$ be equipped with an adjoint-invariant metric $\langle \bullet , \bullet \rangle$, such that the operator $R$ is skew-symmetric: $\langle R\circ a, b\rangle =-\langle a, R\circ b\rangle$. Most importantly, $R$ is required to satisfy the following equation:
\bea\label{Nij}
N(a, b):=[R\circ a, R\circ b]-R\circ ([R\circ a, b]+[a, R\circ b])-[a, b]=0\quad\quad\forall\, a, b \in \mathfrak{g}
\eea
Below we will discuss the meaning of this equation, which often goes in the literature under the clumsy name of `modified classical Yang-Baxter equation'. For the moment we simply state that, once an operator $R$ with these properties is given, the action of the deformed model, expressed in terms of the light-cone components $J_\pm$ of the usual current $J:=-g^{-1}dg$, is:
\bea\label{action}
\mathscr{S}_\eta=\frac{1}{2}\,\int\;d^2x\;\langle J_+, \frac{1+\eta^2}{1-\eta\, R}\circ J_-\rangle,
\eea
where $\eta \in \CC$ is a deformation parameter. Note that we have included in the definition a factor of $(1+\eta^2)$, to facilitate the analysis of certain limits below (Section \ref{limitcase}). The easiest way of constructing the Lax pair for this model is by showing that the Noether current of the model is in fact flat (which in itself is a rather nontrivial property). We will define the latter as follows:
\bear\label{Noether}
&\mathcal{K}_+=-g\,\left( \frac{1+\eta^2}{1+\eta\, R}\circ J_+\right) \,g^{-1}:=-(1+\eta^2)\,g S_+ g^{-1},& \\ &\mathcal{K}_-=-g\,\left( \frac{1+\eta^2}{1-\eta\, R}\circ J_-\right) \,g^{-1}:=-(1+\eta^2)\,g S_- g^{-1}&
\eear
Now, the equation of Noether current conservation $\dd_+\mathcal{K}_-+\dd_-\mathcal{K}_+=0$ can be written, in terms of $S$, as follows:
\bea\label{eom1}
\dd_+S_-+\dd_-S_++\eta\,[R\circ S_-, S_+]-\eta\, [R\circ S_+, S_-]=0
\eea
(We have substituted $S$ in place of $J$ using the definition ${J_\pm=(1\pm \eta R)\circ S_\pm}$.) The flatness condition for the current $J$ (which follows from the definition $J=-g^{-1}dg$) can be also recast in terms of $S$. Using, in particular, the crucial relation (\ref{Nij}), we arrive at the following expression:
\bear
\dd_+S_--\dd_-S_++\eta\,[S_+, R\circ S_-]-\eta\, [R\circ S_+, S_-]-(1-\eta^2)\,[S_+, S_-]-\\ \nonumber
-\eta\, R\circ\left(\dd_+ S_-+\dd_- S_++\eta\,[R\circ S_-, S_+]-\eta\, [R\circ S_+, S_-]\right)=0
\eear
The second line is zero due to the equation of motion (\ref{eom1}), and one can check that the vanishing of the first line is precisely the condition for the flatness of the Noether current: $d\mathcal{K}-\mathcal{K}\wedge\mathcal{K}=0$. Therefore we have constructed a current $\mathcal{K}$, which is both conserved and flat. This is a situation, to which the canonical construction of Pohlmeyer \cite{Pohlmeyer} is applicable. One can build a one-parametric family of flat connections $\mathcal{A}$ as follows:
\bea
\mathcal{A}={1+v\over 2}\,\mathcal{K}_+\,dx^++{1+v^{-1}\over 2}\,\mathcal{K}_-\,dx^-\;, \;\;v\in \CC^\ast\,.
\eea
The zero-curvature equation reads
\bea
\dd_+ \mathcal{A}_--\dd_- \mathcal{A}_+-[\mathcal{A}_+, \mathcal{A}_-]=0\quad\quad \textrm{for all}\;\;v\in \CC^\ast\;.
\eea
Before proceeding further with the discussion of the geometric meaning of the $R$-matrix, for completeness we will describe the generalizations of the above construction to the deformations of symmetric space $\sigma$-models, as well as to two-parametric deformations of the principal chiral model.

\subsection{Symmetric spaces.}\label{symmspacesec}

In order to pass from target space $G$ to its quotient $G/H$, one can `gauge' the subgroup $H$ of the symmetry group of the theory. The symmetry of the undeformed theory, which is $G=G_L\times G_R$, is clearly deformed by the $R$-matrix, and what remains is $G_L\times \widetilde{H}$, where $\widetilde{H}\subset G_R$ is the subgroup commuting with $R$. Let us consider two cases:

\vspace{0.3cm}
\noindent$\bullet$\quad Suppose $H\subset \widetilde{H}$. This means that $R$ commutes with the adjoint action of $H$, i.e. $R(hah^{-1})=h R(a) h^{-1}, h\in H$. This is a very stringent restriction. For example, consider the case that the symmetric space is a sphere $S^N=SO(N+1)/SO(N)$. Decompose $so(N+1)=so(N)\oplus \mathfrak{m}$, then $\mathfrak{m}$ is irreducible over $\CC$ as a representation of $so(N)$. The adjoint representation of $so(N)$ is also irreducible at least for $N>4$. By Schur's lemma, the restrictions of $R$ to $so(N)$ and to $\mathfrak{m}$ are multiples of the identity: $R|_{so(N)}=s \,\mathds{1}$, $R|_{\mathfrak{m}}=r \,\mathds{1}$. The equation (\ref{Nij}) then implies that $R=i\,\mathds{1}$ on the whole algebra. The action (\ref{action}) in this case contains no deformation. The argument does not rule out more sophisticated cases, where the representations of the stabilizer are reducible, however we will not develop this line further here.  

\vspace{0.3cm}
\noindent$\bullet$\quad A universal tool arises from gauging a subgroup $H\subset G_L$ of the left-symmetry group. This amounts to changing, in the action (\ref{action}), ordinary derivatives to covariant derivatives, i.e. the gauged model is defined by the new action
\bea\label{actiongauged}
\mathcal{S}_\eta=\int\;d^2x\;\langle J_+, \frac{1}{1-\eta\, R}\circ J_-\rangle,\;\;J=-g^{-1}(d-A)g,\quad A\in \mathfrak{h}.
\eea
This breaks the global symmetry $G_L$ down to the quotient $N_{G_L}(H)/H$ of the normalizer of $H$ in $G_L$ by $H$. Indeed, for an element $k\in N_{G_L}(H)$ the transformation $g\to k g$ may be compensated by an additional transformation $A\to k A k^{-1}$. The latter is an allowed transformation precisely because, due to the definition of normalizer, $k A k^{-1} \in \mathfrak{h}$. Clearly, the model defined by (\ref{actiongauged}) has target space $G/H$. However, a nontrivial question is whether the gauging just introduced preserves the integrable structure, i.e. if one still has a zero-curvature representation for the e.o.m. Let us check this. The e.o.m. following from (\ref{actiongauged}) comprise two different equations -- one coming from the variation of the group element $g$, the other from the variation of the auxiliary field $A$:
\bear\label{covconseq}
&&\mathscr{D}_+ \mathds{K}_-+\mathscr{D}_- \mathds{K}_+=0,\quad\quad \\ \label{nohcomp}
&& [\mathds{K}_-]_{\mathfrak{h}}=[\mathds{K}_+]_{\mathfrak{h}}=0
\eear
The covariant derivative is $\mathscr{D}M:=dM-[A, M]$, and $\mathds{K}$ is the `current' defined as follows:
\bea\label{currK}
\mathds{K}_-={1\over 1-\eta R_g}\circ (D_-g g^{-1}),\quad \mathds{K}_+={1\over 1+\eta R_g}\circ (D_+g g^{-1}),\quad R_g=Ad_g R Ad_{g^{-1}}\,.
\eea
In the absence of gauging $\mathds{K}$ would be the conserved Noether current, however after gauging most (or all) of the left symmetry will be lost, so this current is only covariantly conserved. Let us now rewrite the definitions (\ref{currK}) of the current as follows:
\bea\label{cartandecomp}
\dd_- g g^{-1}=A_-+(1-\eta R_g)\circ \mathds{K}_-,\quad\quad \dd_+ g g^{-1}=A_++(1+\eta R_g)\circ \mathds{K}_+\,.
\eea
Recalling now eq. (\ref{nohcomp}), i.e. $[\mathds{K}_-]_{\mathfrak{h}}=[\mathds{K}_+]_{\mathfrak{h}}=0$, as well as the fact that $[A_-]_{\mathfrak{m}}=[A_+]_{\mathfrak{m}}=0$, we see that (\ref{cartandecomp}) resembles the decomposition of $dg g^{-1}$ into a `vielbein' and `connection' parts. It allows to calculate $A$ and $\mathds{K}$ in terms of the dynamical group element $g$. One way to do it is to use perturbation theory in $\eta$. At $\eta=0$ one simply has $A=[dg g^{-1}]_{\mathfrak{h}}$ and $\mathds{K}=[dg g^{-1}]_{\mathfrak{m}}$.

The flatness of $dg g^{-1}$ implies a deformed version of Cartan's structure equations:
\bea\label{cartaneq}
dA-A\wedge A=(1+\eta^2) \mathds{K}\wedge \mathds{K}-\mathscr{D} \mathds{K}+\eta \,R_g\circ (\mathscr{D} \ast \mathds{K})
\eea
Let us now use the equation of motion (\ref{covconseq}) to eliminate the last term in the structure equation, and project the remaining equation onto its $\mathfrak{h}$- and $\mathfrak{m}$-components:
\bear
&&\mathfrak{h}: \quad\quad dA-A\wedge A-(1+\eta^2) \mathds{K}\wedge \mathds{K}=0\\
&&\mathfrak{m}:\quad\quad \mathscr{D} \mathds{K}=0
\eear
We have made use of the symmetric space commutation relations: $[\mathfrak{h}, \mathfrak{h}]\subset \mathfrak{h}$, $[\mathfrak{h}, \mathfrak{m}]\subset \mathfrak{m}$ and $[\mathfrak{m}, \mathfrak{m}]\subset \mathfrak{h}$. The above two equations, together with $\mathscr{D} \ast \mathds{K}=0$, may be seen to arise from the zero-curvature condition for the Lax connection
\bea
\mathcal{A}=A+\sqrt{1+\eta^2}\,\left(u \,\mathds{K}_+ dx^++{1\over u}\, \mathds{K}_- dx^-\right)\,.
\eea
We note that the original proof of the integrability of the $\eta$-deformed model for symmetric target spaces was given in \cite{VicedoSymm}.

\subsection{The case of non-simple symmetry group:\\Two-parametric deformations.}\label{2param}

As mentioned earlier, the group $G$ itself may be thought of as a symmetric space: $G=\frac{G\times G}{G_{\mathrm{diag}}}$. The corresponding `gauged $\sigma$-model' Lagrangian is
\bea
\mathscr{L}=\mathrm{Tr}(g_1^{-1}\mathscr{D}g_1)^2+\mathrm{Tr}(g_2^{-1}\mathscr{D}g_2)^2\,.
\eea
$g_1$ and $g_2$ are two group elements, and $\mathscr{D}g:=dg-A\,g, A\in \mathfrak{g}$. We can now deform this action similarly to the way it is done in (\ref{action}), but now with two parameters:
\bea\label{lagrgauged1}
\widetilde{\mathscr{L}}=\mathrm{Tr}\left((g_1^{-1}\mathscr{D}_+g_1)\frac{1}{1-\eta_1 R_1}\circ(g_1^{-1}\mathscr{D}_-g_1)\right)+\mathrm{Tr}\left((g_2^{-1}\mathscr{D}_+g_2)\frac{1}{1-\eta_2  R_2}\circ(g_2^{-1}\mathscr{D}_-g_2)\right)\,.
\eea
We will now integrate out the auxiliary gauge field $A=(A_+, A_-)$. The procedure is described in Appendix \ref{integrout}, and the result is:
\bea\label{gaugedfinlagr1}
\widetilde{\mathscr{L}}={1\over 2}\mathrm{Tr}\left((J_1-J_2)_+ \frac{1}{1-{\eta_1\over 2} R_{g_1}-{\eta_2\over 2} R_{g_2}} \circ (J_1-J_2)_-\right),\quad\quad R_g:=Ad_g\,R\,Ad_{g^{-1}}\,.
\eea
Clearly, this is still gauge-invariant with respect to the transformations $g_1\to g_1 g$, $g_2\to g_2 g,\;g\in G$. The simplest gauge condition would be to set $g_1=1$ or $g_2=1$.

Let us now obtain a zero-curvature representation for the e.o.m. arising from the Lagrangian (\ref{lagrgauged1}). Its derivation is rather similar to the analogous construction for the deformed symmetric-space $\sigma$-models considered above, and the details are discussed in Appendix \ref{laxpair}. Introducing the current
\bea
(\mathscr{K})_\pm={1\over 1\pm\eta_1 R_{g_1}}\circ (\mathscr{D}_\pm g_1 g_1^{-1}),
\eea
we can write the one-parametric family of flat connections as follows:
\bear\nonumber
&\!\!\!\!\!\!\!\mathscr{A}=A+\left(ab+\sqrt{(1+a^2)(1+b^2)}\, u\right)\,\mathscr{K}_+ dx^++\left(ab+\sqrt{(1+a^2)(1+b^2)}\, {1\over u}\right)\,\mathscr{K}_- dx^-,&\\ \nonumber
&a={\eta_1+\eta_2\over 2}, \quad\quad b={\eta_1-\eta_2\over 2},\quad\quad u\in \CC^\ast\,.&
\eear
The two-parametric model we are considering was originally introduced in the form (\ref{gaugedfinlagr1}) in \cite{Klimcik} and its Lax pair was found in \cite{Klimcik2}. The analysis presented here is closest to the one of \cite{Hoare}.

\section{\Large What can the $R$-matrix be? \\Relation to complex structures.}\label{Rmatrix}

Let us now turn to the analysis (and interpretation) of the principal equation~(\ref{Nij}). At first glance one notices that, were $R$ a complex structure, i.e. if $R^2=-1$, equation (\ref{Nij}) would have been identical to the equation of integrability of the complex structure $R$ on the Lie group $G$ with Lie algebra $\mathfrak{g}$. (The l.h.s. of (\ref{Nij}) is simply the Nijenhuis tensor evaluated on two vectors $a, b$.) Therefore we already know a whole class of solutions of (\ref{Asymm})-(\ref{Nij}): these are given by complex structures on the Lie group $G$, compatible with the metric $\langle\bullet, \bullet \rangle$ (also called \emph{orthogonal complex structures}). Compatibility with the metric means here that the metric is Hermitian w.r.t. the chosen complex structure $R$, i.e. $\langle R\circ a, R \circ b\rangle=\langle a, b\rangle$. Since $R^2=-1$, this is the same as the skew-symmetry of $R$, as defined in Section \ref{general}.

The observation that properties of the $R$-matrix are akin to those of a complex structure will be crucial for the rest of our discussion, however at present we will analyze the limitations of this parallel. First of all, the existence of  complex structures on a Lie group crucially depends on its dimension, or in particular whether it is even or odd. Obviously, an odd-dimensional Lie group cannot admit a complex structure. Many important examples of target spaces for $\sigma$-models, such as $S^3=SU(2)$, fall in this category. In this situation one possibility, which has been elaborated in the literature \cite{Klimcik}-\cite{Vicedo}, is to choose a matrix $R$, satisfying (\ref{Asymm})-(\ref{Nij}) and an additional requirement
\bea
R^3=-R
\eea
in place of $R^2=-1$ that we have in the case of complex structure. An operator $R$ satisfying this relation is sometimes called an $f$-structure on the manifold. This operator can have, on $\mathfrak{g}_\CC$, the eigenvalues $0, \pm i$. Let us decompose $\mathfrak{g}_\CC$ into the corresponding eigenspaces:
\bea
\mathfrak{g}_\CC=\mathfrak{m}_0\oplus \mathfrak{m}_+\oplus\mathfrak{m}_-\;.
\eea
What does the vanishing of the tensor $N$ imply for the commutation relations between the elements of these subspaces? One has the following result:
\bear
a \in \mathfrak{m_+}, \;\;b\in\mathfrak{m_-}:\quad\quad &&N(a, b)\equiv 0\\ 
\label{Nij2}
a\in \mathfrak{m}_+, \;\;b\in\mathfrak{m}_+: \quad\quad &&N(a, b)=-2i\, R\circ[a, b]-2[a, b], \\ \label{Nij3}
a\in \mathfrak{m}_-, \;\;b\in\mathfrak{m}_-: \quad\quad &&N(a, b)=2i\, R\circ[a, b]-2[a, b], \\
a\in \mathfrak{m}_0, \;\;b\in\mathfrak{m}_+: \quad\quad &&N(a, b)=-i\,R\circ[a, b]-[a,b], \\
a\in \mathfrak{m}_0, \;\;b\in\mathfrak{m}_-: \quad\quad &&N(a, b)=i\,R\circ[a, b]-[a,b], \\
a\in \mathfrak{m}_0,\, \;\;b\in\mathfrak{m}_0\,: \quad\quad &&N(a, b)=-[a,b] \,.
\eear
The vanishing of $N$ leads to the following commutation relations:
\bea\label{commrel}
[\mathfrak{m}_\pm, \mathfrak{m}_\pm]\subset \mathfrak{m}_\pm,\quad [\mathfrak{m}_0, \mathfrak{m}_\pm]\subset \mathfrak{m}_\pm,\quad [\mathfrak{m}_0, \mathfrak{m}_0]=0\;.
\eea
Let us use the standard decomposition
\bea\label{mpm}
\mathfrak{g}_\CC=\mathfrak{t}\oplus_{\alpha>0} \mathfrak{g}_\alpha \oplus_{\alpha>0} \mathfrak{g}_{-\alpha}
\eea
of the Lie algebra into the Cartan subalgebra and the positive and negative root subspaces. One can choose the action of $R$ in the following way:
\bear\label{Rroot1}
&&R\big|_{\mathfrak{g}_\alpha}=i\;\mathds{1},\quad \alpha>0\\ \label{Rroot2}
&&R\big|_{\mathfrak{g}_\alpha}=-i\;\mathds{1},\quad \alpha<0\\ \label{Rroot3}
&&R\big|_{\mathfrak{t}}=0\,.
\eear
Clearly, this choice is compatible with the commutation relations above. There is, however, a subtlety which concerns the reality of the action (\ref{action}). Indeed, in (\ref{Rroot1})-(\ref{Rroot3}) we have defined the operator $R$ on the \emph{complexification} $\mathfrak{g}_\CC$ of the original Lie algebra~$\mathfrak{g}$. There is in general no reason why $R$, defined as above, should restrict to the real form $\mathfrak{g}$.  In other words, let us view $\mathfrak{g}$ as the set of fixed points of a semi-linear automorphism $^\ast: \mathfrak{g}_\CC \to \mathfrak{g}_\CC$ (`complex conjugation'). In the action (\ref{action}) we assume $^\ast(J_\pm)=J_\pm$ (for Minkowskian signature on the worldsheet, which we have been using so far). Therefore reality of the action will depend on the commutation relation between $R$ and $^\ast$, as well as on the complex phase $\phi$ of the parameter $\eta=|\eta|e^{i \phi}$. The most general condition, for which the Lagrangian is real in Minkowskian signature (up to an overall factor of $1+\eta^2$), is
\bea
^\ast R= e^{2i \phi} R\,^\ast 
\eea
Below we will be interested in the case when the operator $R$, defined by (\ref{Rroot1})-(\ref{Rroot3}), descends to $\mathfrak{g}$, i.e. it commutes with $^\ast$, hence ${\phi=0 \,(\mathrm{mod}\;\pi)}$. For example, for ${\mathfrak{g}=su(N)}$ this is so when $^\ast(a):=-a^\dagger$ (here $a$ is viewed as an $N\times N$ matrix) -- then  $\mathfrak{g}$ is a compact real form of $\mathfrak{g}_\CC$. In our applications we will have  $\mathfrak{g}_\CC=sl(2, \CC)$, $\mathfrak{g}=su(2)$. Another situation that we will encounter is when $^\ast(a):=-\sigma_3 a^\dagger\sigma_3^{-1}$. The set of fixed points is the Lie algebra $su(1,1)$. In both examples the positive/negative roots can be thought of as the matrices $\left( \begin{array}{cc}
0 & 1   \\
0 & 0  
 \end{array} \right), \left( \begin{array}{cc}
0 & 0   \\
1 & 0  
 \end{array} \right)$.
 
In general the reality properties of $R$ depend not only on the real form, but also on the particular choice of root space decomposition. For example, $su(1,1)$ is equivalent to $sl(2, \mathbb{R})$\footnote{Two real forms of $sl(2,\CC)$, defined by the conjugation operators $^{\ast_1}, ^{\ast_2}$ are said to be equivalent if $^{\ast_2}=Ad_g \;^{\ast_1} Ad_{g^{-1}}$ for some $g\in SL(2,\CC)$. For the case at hand $^{\ast_1}(a)=\bar{a}, \;^{\ast_2}(a)=-\sigma_3 a^\dagger\sigma_3^{-1}$ and the conjugating element $g$ is given by $g={1\over \sqrt{2}}\left( \begin{array}{cc}
1 & i   \\
i & 1  
 \end{array} \right)$.}. The latter real form is defined by $^\ast(a)=\bar{a}$ (usual complex conjugation). If one chooses the positive/negative roots to be the same upper/lower-triangular matrices as above \emph{in this basis}, the operator $R$ will be purely imaginary, i.e. it will anti-commute with $^\ast$, hence for the Lagrangian to be real one would need to take ${\phi={\pi\over 2}\, (\mathrm{mod}\;\pi)}$. 

\subsection{The limit $\eta\to\pm i$.}\label{limitcase}

So far we have been focusing primarily on Minkowskian worldsheets. To make contact with the models of \cite{Bykov3}, let us consider the Euclidean case as well. We then have to replace $x^+\to z, x^- \to \bar{z}$, and the conjugation operator now permutes the currents: $^\ast(J_z)=J_{\bar{z}}$. Using the antisymmetry of $R$, we see that the reality condition is now $^\ast R= -e^{2i \phi} R\,^\ast $, so that $\eta$ has to be multiplied by an additional factor of $i$ compared to the Minkowskian case.

In order for this analysis not to be void of examples, let us consider the following application. We will consider a compact group $G$ and a Euclidean worldsheet. For the action to be real, we will consider pure imaginary values of $\eta$. Let us now rewrite the action (\ref{action}) as follows:
\bea
\mathscr{S}_\eta={1\over 2}\,\int\;d^2x\;\left((1+i\,\eta)\,\langle J_z, (J_{\bar{z}})_{\mathfrak{m}_+}\rangle+(1-i\,\eta)\,\langle J_z,  (J_{\bar{z}})_{\mathfrak{m}_-}\rangle+(1+\eta^2)\,\langle J_z, (J_{\bar{z}})_{\mathfrak{m}_0}\rangle\right)
\eea
We may notice now that there exist limits $\underset{\eta\to \pm i}{\mathrm{lim}}\, \mathscr{S}_\eta$. For definitiveness, let us consider the limit $\eta \to i$. In this case
\bea\label{Slim}
\mathscr{S}_{\mathrm{lim}}=\underset{\eta\to i}{\mathrm{lim}}\,\mathscr{S}_{\eta}= \int\;d^2x\;\langle (J_z)_{\mathfrak{m}_+},  (J_{\bar{z}})_{\mathfrak{m}_-}\rangle
\eea
(We have appealed to the property $\langle\mathfrak{m}_0, \mathfrak{m}_-\rangle=\langle\mathfrak{m}_-, \mathfrak{m}_-\rangle=0$, which follows from the antisymmetry of $R$.)
The Noether current (\ref{Noether}) and hence the Lax pair also have well-defined limits. Amusingly, the target space of the limiting model is different from the original one. The reason for this is that the action $\mathscr{S}_{lim}$ is invariant under a group $H$ of gauge transformations, where $H$ is the group with Lie algebra $\mathfrak{m}_0$. The $\mathfrak{m}_\pm$-projections of the  currents are tensors under this group, i.e. they transform as $(J)_{\mathfrak{m}_\pm}\to h(J)_{\mathfrak{m}_\pm} h^{-1}$, and the action above is invariant. As a result, the target space is $G/H$ and not $G$, as it was for the original model (\ref{action}). A particularly transparent example is $G=SU(N)$, $H=U(1)^{N-1}$ (the Cartan torus of $G$). In this case the target space of the limiting model is $G/H=SU(N)/U(1)^{N-1}$ -- the manifold of complete flags in $\CC^N$. For $N=3$ this is the model discussed in \cite{Bykov1}-\cite{Bykov2}.

To finalize this discussion, let us point out that, although the models of \cite{Bykov1}-\cite{Bykov2} are obtained from (\ref{action}) as particular limits, some generalizations thereof considered in~\cite{Bykov3} cannot be obtained in this way. In \cite{Bykov3} it was pointed out that an action of the type (\ref{Slim}) defines a model with a Lax connection for any \emph{complex homogeneous} target space, whose Killing metric is compatible with the complex structure. As an example, one can consider partial flag manifolds $SU(N)/S(U(n_1)\times \ldots \times U(n_m))$, $\sum\limits_{k=1}^m n_k=N$\footnote{For target spaces of this type, an alternative construction of Lax pairs was carried out in \cite{Young}, using the $\mathbb{Z}_m$-graded structure of such spaces. The equivalence of the actions of the two models (the one in \cite{Young} and (\ref{Slim})) was proven in \cite{Bykov4}. A natural expectation is that the Lax connections produced by the two constructions are gauge-equivalent. }. On the other hand, the limiting procedure described above always gives target spaces of the form $G/H$, where $H$ is an \emph{abelian} subgroup of $G$. This follows from the fact that its Lie algebra $\mathfrak{m}_0$ is abelian, according to (\ref{commrel}).

\subsection{Complex structures on Lie groups}\label{complstruct}
Let us return to a different choice of $R$: henceforth we will choose $R$ to coincide with a complex structure on a Lie group $G$, as described at the beginning of Section \ref{Rmatrix}. First of all, one has the following result:

\vspace{0.3cm}
\textbf{Statement 1.} On an even-dimensional compact simple Lie group there always exists an orthogonal complex structure. (Here we mean orthogonality w.r.t. the Killing metric).

\vspace{0.3cm}
\textbf{Proof.} The tangent space to a Lie group $G$ at a point $g\in G$ is spanned by the vielbein $g^{-1}dg\in \mathfrak{g}$. Therefore specifying a left-invariant almost complex structure on a compact Lie group is the same as specifying a complex structure on the Lie algebra $\mathfrak{g}$. If the latter is of even dimension, this is certainly always possible. Therefore the question is whether it can be chosen to be orthogonal and integrable. We will be denoting the complex structure on the Lie algebra and the respective almost complex structure on the Lie group by the same letter $\mathscr{J}$. Let us assume that a complex structure $\mathscr{J}$ on $\mathfrak{g}$ has been chosen, i.e.
\bea
\mathscr{J}:\quad \mathfrak{g} \to \mathfrak{g},\quad\quad \mathscr{J}^2=-\mathds{1}.
\eea
Requiring integrability of $\mathscr{J}$, viewed as a complex structure on $G$, is a rather stringent condition. To this end, let us diagonalize $\mathscr{J}$ in $\gfr_\CC$ and denote its holomorphic and anti-holomorphic eigenspaces by $\mg_\pm\subset \gfr_\CC$:
\bea
\mathscr{J}\circ \mg_\pm=\pm i\;\mg_\pm\;.
\eea
Clearly, $\gfr_\CC=\mg_+\oplus\mg_-$. Considering the value of the Nijenhuis tensor $N(a, b):=[R\circ a, R\circ b]-R\circ ([R\circ a, b]+[A, R\circ b])-[a, b]$ on the elements of the subspaces $\mathfrak{m}_\pm$, we get:
\bear
a \in \mathfrak{m_\pm}, \;\;b\in\mathfrak{m_\mp}:\quad\quad &&N(a, b)\equiv 0, \\ \label{Nij2}
a\in \mathfrak{m}_+, \;\;b\in\mathfrak{m}_+:\quad\quad &&N(a, b)=-2i\, R\circ[a,b]-2[a,b], \\ \label{Nij3}
a\in \mathfrak{m}_-, \;\;b\in\mathfrak{m}_-: \quad\quad &&N(a, b)=2i\, R\circ[a,b]-2[a,b] \,.
\eear
Therefore the vanishing of $N(a, b)$ is equivalent to the requirement that $\mg_\pm$ be subalgebras of $\gfr_\CC$:
\bea\label{commcomplstruct}
[\mg_+, \mg_+]\subset \mg_+,\quad\quad [\mg_-, \mg_-]\subset \mg_-\;.
\eea
Just as before, we will use the standard decomposition
\bea\label{mpm}
\mathfrak{g}_\CC=\mathfrak{t}\oplus_{\alpha>0} \mathfrak{g}_\alpha \oplus_{\alpha>0} \mathfrak{g}_{-\alpha}
\eea
of the Lie algebra into the Cartan subalgebra and the positive and negative root subspaces. Since all roots are paired, one sees that $\dim\,\mathfrak{g} = \dim \mathfrak{t}\;(\mathrm{mod}\;2)$. Let us now choose the action of the complex structure $\mathscr{J}$ on $\mathfrak{g}_{\pm\alpha}$ as follows:
\bear\label{Jborel1}
\mathscr{J}(a)=i\,a\quad \mathrm{for}\quad a\in \mathfrak{g}_\alpha \;\;(\alpha>0)\\ \label{Jborel2}
\mathscr{J}(a)=-i\,a\quad \mathrm{for}\quad a\in \mathfrak{g}_\alpha \;\;(\alpha<0)
\eear
In other words, $\oplus_{\alpha>0} \mathfrak{g}_\alpha \subset \mathfrak{m}_+$ and $\oplus_{\alpha>0} \mathfrak{g}_{-\alpha}\subset \mathfrak{m}_-$. This is clearly compatible with the commutation relations (\ref{commcomplstruct}). So far we have essentially repeated the definitions (\ref{Rroot1}), (\ref{Rroot2}). Orthogonality of $\mathscr{J}$ w.r.t. to the Killing metric, i.e. Hermiticity of the latter, is equivalent to the statement that the $(0, 2)$ and $(2, 0)$ components of the metric vanish: $\langle \mathfrak{m}_\pm, \mathfrak{m}_\pm \rangle=0$. This is satisfied by the above choice (\ref{Jborel1})-(\ref{Jborel2}), as is obvious from the standard commutation relations of the elements of the Cartan subalgebra and of the root subspaces and the definition of the Killing metric $\langle a, b\rangle:=\mathrm{Tr}(\mathrm{ad}_a \mathrm{ad}_b)$. What remains is to extend the above definitions to elements of the Cartan subalgebra $\mathfrak{t}$, which is even-dimensional due to the even dimensionality of the group. Since $[\mathfrak{t}, \mathfrak{g}_\alpha]\subset \mathfrak{g}_\alpha$, the vanishing of the Nihenjuis tensor (\ref{Nij2})-(\ref{Nij3}) does not impose any additional constraints. Therefore we may choose an arbitrary complex structure on $\mathfrak{t}$, as long as it is compatible with the metric. In the basis, in which the restriction of the metric $\langle\bullet, \bullet \rangle\big|_\mathfrak{t}$ to $\mathfrak{t}$ is proportional to the unit matrix, this simply means we need to choose a \emph{skew}-symmetric matrix $\mathscr{J}$, which squares to minus the identity $\mathscr{J}^2=-1$. Clearly, this is always possible. $\blacksquare$

\vspace{0.3cm}
\textbf{Comment.} The statement of the theorem will also hold if one replaces the simple group by a semi-simple group times a torus (i.e. a reductive group), provided one extends the metric to the toric part of the group in a non-degenerate way. Below we will be considering a particular example of such situation, namely the group $U(2)=U(1)\times SU(2)$.

\vspace{0.3cm}
A Lie group with at least one integrable left-invariant complex structure will generally possess a continuous family of integrable complex structures (the right shifts of the original complex structure). An example of this phenomenon is furnished by the so-called hypercomplex groups, which are particular cases of hypercomplex manifolds. A hypercomplex manifold is a manifold with a triple $(\mathscr{I}, \mathscr{J}, \mathscr{K})$ of pairwise anti-commuting integrable complex structures satisfying $\mathscr{I}\circ \mathscr{J}=\mathscr{K}$. Such manifolds are important as they are examples of reduced holonomy manifolds -- the unique torsion-free affine connection on such a manifold of dimension $4n$ (called `Obata connection') has holonomy contained in $GL(n, \mathds{H})$. One may think of these manifolds as generalizations of hyper-K\"ahler manifolds, when one relaxes the requirement that the covariant derivative preserves the metric. Hypercomplex groups were found and classified in \cite{Troost} and \cite{Joyce}. One particularly interesting example in low dimensions is $G=U(2)=S^1\times S^3$, which we will focus on below. We will show that left-invariant complex structures on $U(2)$ are in one-to-one correspondence with complex structures on $\mathbb{R}^4$. Therefore we come to a review of the latter.

\section{\Large The $S^1\times S^3$ model}\label{S1S3}
\subsection{Complex structures on $\mathbb{R}^4$}\label{R4complstruct}

A complex structure on $\mathbb{R}^4$ is an operator $\mathscr{J}: \mathbb{R}^4 \to \mathbb{R}^4$, satisfying
\bea\label{cs}
\mathscr{J}^2=-\mathds{1}_4.
\eea
An \emph{orthogonal} complex structure is one compatible with the Euclidean metric on $\mathbb{R}^4$, i.e. $\mathscr{J}^t \mathscr{J}=\mathds{1}_4$. Together with (\ref{cs}), this implies $\mathscr{J}^t=-\mathscr{J}$. Therefore the space of orthogonal complex structures on $\mathbb{R}^4$ is the space of skew-symmetric $4\times 4$ matrices satisfying (\ref{cs}).

It follows from (\ref{cs}) that $\det \mathscr{J}=\pm 1$. Since the determinant of a real skew-symmetric matrix is non-negative, $\det \mathscr{J}=1$. On the space $\bigwedge^2 \mathbb{R}^4$ of skew-symmetric matrices one can define the duality operator $\ast$ in the usual way:
\bea
\ast:\quad \bigwedge\!^2\; \mathbb{R}^4 \to \bigwedge\!^2\;\mathbb{R}^4,\quad (\ast \mathscr{J})_{ab}:={1\over 2}\epsilon_{abcd} \mathscr{J}_{cd}\,.
\eea
It is an involution, i.e. $\ast^2=1$. We will now prove that an orthogonal complex structure is necessarily self- or anti-self-dual: $\ast \mathscr{J}=\pm \mathscr{J}$, depending on the value of the Pfaffian\footnote{We recall the definition of the Pfaffian: $\mathrm{Pf}(\mathscr{J}):={1\over 8}\,\epsilon_{abcd}\,\mathscr{J}_{ab}\,\mathscr{J}_{cd}$.} $\mathrm{Pf}(\mathscr{J})=\pm 1$. To this end, consider the quantity
\bea
\tr(\mathscr{J}\mp \ast \mathscr{J})^2=\tr \mathscr{J}^2+\tr(\ast \mathscr{J})^2\mp 2\,\tr(\mathscr{J} \ast \mathscr{J})
\eea
It is easy to show that $\tr(\ast \mathscr{J})^2=\tr(\mathscr{J}^2)=-4$ and $\tr(\mathscr{J} \ast \mathscr{J})=-4 \mathrm{Pf}(\mathscr{J})=\mp 4$. Substituting in the above formula, we obtain
\bea
\tr(\mathscr{J}\mp \ast \mathscr{J})^2=-4-4+8=0.
\eea
For a skew-symmetric matrix $C$ the property $\tr (C^2)=0$ implies $C=0$, hence
\bea
\mathscr{J}=\pm\ast \mathscr{J}\,.
\eea

\subsection{$S^1\times S^3$ as a hyper-hermitian manifold}\label{hyperhermit}

Let us pick a basis of anti-Hermitian generators of the Lie algebra $\mathfrak{u}(2)$, ${\sigma_\mu=i\,\{\mathbf{1}, \vec{\sigma}\},}$ $\mu=0, 1, 2, 3$. Given a complex structure $\mathscr{J}$ on $\mathfrak{u}(2)=\mathbb{R}^4$, the corresponding holomorphic subspace is:
\bea\label{holspaces}
\mg_+=\mathrm{Span}\left((\sigma_+)_\mu=\sum\limits_{\nu=0}^3\,\left(\mathds{1}-i\, \mathscr{J}\right)_{\mu\nu} \sigma_\nu,\quad \mu=0, 1, 2, 3\right)
\eea
The fact that this complex structure is integrable is not obvious and requires a special check. We conduct such check in Appendix \ref{integrcomplstruct}. In fact, we can deduce additional information from the discussion there. Define the scalar product $\langle a, b\rangle$ on the Lie algebra $\mathfrak{u}(2)$ as follows:
\bea
\langle a, b\rangle=\tr(ab)\,.
\eea
Here $a, b$ are taken as $2\times 2$ anti-Hermitian matrices, just like the $\sigma_\mu$ above.
This is a non-degenerate $ad$-invariant scalar product, generalizing the Killing product (which in this case is degenerate). The results of Appendix \ref{integrcomplstruct} are summarized by the following two statements:

\vspace{0.3cm}
\textbf{Lemma 1.} The space of integrable (left-)invariant complex structures on $S^1\times S^3$ coincides with the space of complex structures on $\mathbb{R}^4$. This space is $\CP^1 \sqcup \CP^1$ -- the disjoint union of two spheres, distinguished by the values of the Pfaffian $\mathrm{Pf}(\mathscr{J})=\pm 1$.

\vspace{0.3cm}
\textbf{Lemma 2.} The metric on $\mathfrak{u}(2)$, defined by $\langle\bullet, \bullet\rangle$, is Hermitian for all complex structures $\mathscr{J}$.

\vspace{0.3cm}
We have thus proven that the manifold $S^1\times S^3$ is hyper-hermitian. This is the crucial property which will allow us to define deformations of the $S^1\times S^3$ principal chiral model below.

\subsection{The Lagrangian}\label{Themodel}

Let $g\in U(2)$ be a group element. Introduce the left-invariant current
$
J:=-g^{-1}dg=J_\mu \sigma_\mu
$. The Lagrangian of the model reads
\bea
\mathscr{L}=(J_\mu)_- \,A_{\mu\nu}\, (J_\nu)_+,\quad\quad A=\mathds{1}+\eta\, \mathscr{J}
\eea
More invariantly, one can write
\bea\label{deflagr}
\mathscr{L}=h(J_-, J_+)+\eta\cdot \,\omega(J_-, J_+),
\eea
where $h$ is the metric on $\mathfrak{u}_2$ introduced earlier, and $\omega$ is the respective K\"ahler form: $\omega=h\circ \mathscr{J}$. This form is not closed: $d\omega\neq 0$. In fact, $S^1\times S^3$ is a prominent example of a complex manifold that is not K\"ahler. The argument goes as follows. If it were K\"ahler, there would exist a closed non-degenerate 2-form $\Omega: d\Omega=0$. The volume would be proportional to $\int \Omega\wedge \Omega$. However, the second cohomology of $S^1\times S^3$ is trivial: $H^2(S^1\times S^3, \mathbb{R})=0$, therefore $\Omega=du$, implying that the volume is zero: $\int \Omega\wedge \Omega=\int \,d(u\wedge \Omega)=0$. The form of the deformed Lagrangian (\ref{deflagr}) is general: whenever one chooses the complex structure $\mathscr{J}$ for the $R$-matrix, one obtains a deformation of the principal chiral model Lagrangian by a term proportional to (the pull-back to the worldsheet of) the respective K\"ahler form on the group target space.

The configuration space of the model is the group $U(2)$ -- a homogeneous space. Moreover, the Lagrangian is invariant under the left action of $U(2)$ (on itself). As a result, the equations of motion are equivalent to the conservation of the corresponding Noether current $\mathcal{K}$:
\bea
\dd_+ \mathcal{K}_-+\dd_- \mathcal{K}_+=0
\eea
One has the following expression for the current:
\bear
\mathcal{K}_+=g \sigma_\mu g^{-1}\,A_{\mu\nu}\, (J_{\nu})_+\\
\mathcal{K}_-=g \sigma_\mu g^{-1}\,A^t_{\mu\nu}\, (J_{\nu})_-
\eear
Introducing the projectors $P_{\mg_\pm}={\mathds{1}\mp i\,\mathscr{J}\over 2}$ on $\mg_\pm$ (such as the one that features in formula (\ref{holspaces})), we can write the matrix $A$ as
\bea
A=(1+i\,\eta)\,P_{\mg_+}+(1-i\,\eta)\,P_{\mg_-}
\eea
Substituting this in the expressions for the components of the current, we get
\bear
\mathcal{K}_+=-g\left((1-i\,\eta)\,(J_+)_{\mg_+}+(1+i\,\eta)\,(J_+)_{\mg_-} \right)g^{-1}\\
\mathcal{K}_-=-g\left((1+i\,\eta)\,(J_-)_{\mg_+}+(1-i\,\eta)\,(J_-)_{\mg_-} \right)g^{-1}
\eear
This model is a particular example of models described in section \ref{general}, where we have specialized the $R$-matrix to be the complex structure $\mathscr{J}$. Therefore the Noether current is flat and leads directly to a one-parametric family of flat connections.

Let us assume for definitiveness that $\ast \mathscr{J}=-\mathscr{J}$ and parametrize the complex structure $\mathscr{J}$ by a unit vector $\vec{n}$ as follows:
\bea\label{Jn}
\mathscr{J}= \left( \begin{array}{cccc}
0 & -n_1 & -n_2 & -n_3 \\
n_1 & 0 & n_3 & -n_2 \\
n_2 & -n_3 & 0 & n_1 \\
n_3 & n_2 & -n_1 & 0
 \end{array} \right)
\eea
Then we can write the Lagrangian explicitly as follows:
\bea\label{Lagr}
\mathscr{L}=(J_0)_+\,(J_0)_{-}+\eta \,(J_0)_- (\vec{n}\cdot \vec{J})_+-\eta \,(J_0)_+(\vec{n}\cdot \vec{J})_- +\vec{J}_+\cdot \vec{J}_{-}+\eta\,\epsilon_{ijk}(J_i)_+(J_j)_{-}\,n_k
\eea
Note that the last term is in fact a full derivative: $\epsilon_{ijk}(J_i)\wedge(J_j)\,n_k\propto d (n_k J_k)$. It is irrelevant for our present purposes, and we will omit it. The limit $\eta\to i$ of this model was considered in \cite{Bykov4}.

\subsubsection{T-duality.}

Let us call $\varphi$ the angle in the $S^1$ direction. In this case $J_0=d\varphi$. It is instructive to see, what T-duality of the above model along the $\varphi$-direction gives. To this end, recall that the target space metric of the model (\ref{Lagr}) is simply a product metric $ds^2=d\varphi^2+\widetilde{ds^2}_{S^3}$. The $B$-field is, up to a total derivative that we have already omitted, $B=\eta\, d\varphi\wedge \vec{n}\vec{J}$. By the Buscher rules \cite{Buscher}, the dual model will have a zero $B$-field and the following metric:
\bea
(ds^2)^{\vee}=(d\widetilde{\varphi}-\eta\,\vec{n}\vec{J})^2 +\widetilde{ds^2}_{S^3}
\eea
The metric on $S^3$ can be written in a form exhibiting the Hopf fibration: \\$\widetilde{ds^2}_{S^3}=\vec{J}\cdot\vec{J}=\vec{J}\cdot \left(1-\vec{n}\otimes \vec{n}\right)\cdot \vec{J}+(\vec{n}\vec{J})^2$. The direction of the Hopf fiber is specified by the vector $\vec{n}$. Calling $\psi$ the angular direction in the Hopf fiber, we may write $\vec{n}\vec{J}=d\psi-A$, where $A$ is the connection on the Hopf bundle over $S^2$. After a linear change of variables, the dual metric takes the form
\bea\label{S1bundle}
(ds^2)^{\vee}=\eta^2\,(d\widehat{\varphi}-A)^2 +(d\psi-A)^2+(ds^2)_{S^2}
\eea
We see that the underlying manifold is the total space of an $S^1\times S^1$ bundle over $S^2$.

\subsubsection{A mechanical reduction.}

Let us pass from light-cone coordinates $x^\pm$ to the usual worldsheet coordinates $(\sigma, \tau)$ by means of the formulae $x^\pm=\sigma\pm\tau$. A natural mechanical reduction of the system (\ref{Lagr}) may be obtained by the following substitution:
\bea
\varphi=
\sigma+\tilde{\varphi}(\tau),\quad\quad (\vec{J})_\tau=(\vec{J})_\tau(\tau),\quad\quad (\vec{J})_\sigma=0
\eea
The last part is the requirement that the $S^3$-angles be independent of $\sigma$. It is easy to convince oneself that this reduction is consistent and leads to a mechanical system with the following Lagrangian:
\bea
\mathcal{L}={1\over 2}\left(\dot{\varphi}^2+(\vec{J})_\tau^2\right)-\eta\,(\vec{n}\vec{J})_\tau\,.
\eea
This Lagrangian describes the motion of a non-relativistic particle on $S^1\times S^3$ interacting with an electromagnetic field on $S^3$, whose potential is $A=\eta\,\vec{n}\vec{J}$. The motion on $S^1$ completely decouples from the motion on $S^3$. The system is, clearly, integrable, as there are four degrees of freedom and four commuting conserved quantities: the energy, momentum in the $\varphi$-direction and two momenta on the three-sphere (one corresponding to left $U(1)$-rotations and the other corresponding to right $U(1)$ rotations leaving vector $\vec{n}$ invariant).

The straightforward mechanical reduction of the T-dual model describes a free particle (geodesic motion) on the manifold with the metric (\ref{S1bundle}).

\section{\Large The $AdS_3\times S^3\times S^1$ model}\label{AdSDef}

In this Section we will consider an application of the above theory in the context of AdS/CFT. In recent years a lot of research concentrated on the study of a Green-Schwarz-type $\sigma$-model with target space $AdS_3\times S^3\times S^3\times S^1$ or $AdS_3\times S^3\times (S^1)^4$ \cite{Maldacena}, \cite{Babichenko}. The latter may be formally thought of as a particular limit of the former, when the radius of one of the spheres diverges. For simplicity we will consider the case of $AdS_3\times S^3\times (S^1)^4$. Technically speaking, the target space of a fully $\kappa$-gauge-fixed Green-Schwarz $\sigma$-model may be viewed as a direct product of $(S^1)^4$ and a homogeneous space
\bea
\mathcal{M}_0=\frac{G_0\times G_0}{H_0},\quad\quad G_0=PSU(1,1|2),\;\; H_0=SU(1,1)\times SU(2)\,,
\eea
whose bosonic part is $AdS_3\times S^3$. Since we have in mind a particular intertwining of $S^3$ and $S^1$ with the help of a complex structure, we will be interested in the manifold
\bea\label{Mmanifold}
\mathcal{M}=\mathcal{M}_0\times S^1=\frac{G\times G}{H},\quad\quad G=G_0\times U(1),\;\; H=H_0\times U(1)\,.
\eea
The remaining three periodic bosons will be simply spectators of the construction to follow. We start with the following lemma:

\vspace{0.3cm}
\textbf{Lemma 3.} Any $R$-matrix on the bosonic subalgebra  $\widehat{\mathfrak{b}}$  of $su(m_1, m_2|n_1, n_2)\oplus u(1)$,
satisfying the Nijenhuis equation (\ref{Nij}) ($N=0$) and the constraint $R^3=-R$,  may be lifted to the full algebra.

\vspace{0.3cm}
\textbf{Proof.} The structure of a $sl(m|n)$ superalgebra is as follows: $sl(m|n)=\mathfrak{b} \oplus \mathfrak{f}_1\oplus \mathfrak{f}_2$, where $\mathfrak{b}$ is a bosonic subalgebra and $\mathfrak{f}_1$, $\mathfrak{f}_2$ are two fermionic $\mathfrak{b}$-modules. These have the following commutation relations:
\bear\label{commrelsuper}
&[\mathfrak{b}, \mathfrak{f}_1]\subset \mathfrak{f}_1,\quad\quad [\mathfrak{b}, \mathfrak{f}_2]\subset \mathfrak{f}_2,&\\ \nonumber
&[\mathfrak{f}_1, \mathfrak{f}_1]=0,\quad\quad [\mathfrak{f}_2, \mathfrak{f}_2]=0,\quad\quad [\mathfrak{f}_1, \mathfrak{f}_2]\subset \mathfrak{b}&
\eear
In the language of Kac \cite{Kac}, $sl(m|n)$ has a $\mathds{Z}$-grading, such that $sl(m|n)_0=\mathfrak{b}$, $sl(m|n)_1=\mathfrak{f}_1$, $sl(m|n)_{-1}=\mathfrak{f}_2$ and $sl(m|n)_i=0$ for $|i|>1$. The commutation relations become transparent if one recalls the `defining' representation of the algebra in terms of $(m+n)\times (m+n)$ matrices: $sl(m|n)=\left( \begin{array}{c|c}
\mathfrak{b} & \mathfrak{f}_1  \\ \hline
\mathfrak{f}_2 & \mathfrak{b}  \end{array} \right)$. Additionally, the $u(1)$ summand commutes with everything else.

Suppose now we have a linear operator $R$ on the bosonic subspace $\widehat{\mathfrak{b}}:=\mathfrak{b}\oplus u(1)$, satisfying eq. (\ref{Nij}). Let us now postulate that
\bea\label{Rferm}
R\big|_{\mathfrak{f}_1}=i\,\mathds{1}, \quad\quad R\big|_{\mathfrak{f}_2}=-i\,\mathds{1}\,.
\eea
It is easy to check that the commutation relations (\ref{commrel}) are still satisfied. Indeed, $\mathfrak{m}_0\subset \widehat{\mathfrak{b}}$, therefore $[\mathfrak{m}_0, \mathfrak{f}_{1,2}]\subset \mathfrak{f}_{1,2}$. Besides, $\mathfrak{m}_+=(\mathfrak{m}_+)_{\widehat{\mathfrak{b}}}\oplus \mathfrak{f}_1$. On the other hand, we know that $[(\mathfrak{m}_+)_{\widehat{\mathfrak{b}}}, \mathfrak{f}_1]\subset \mathfrak{f}_1$ and $[\mathfrak{f}_1, \mathfrak{f}_1]=0$, hence $[\mathfrak{m}_+, \mathfrak{m}_+]\subset \mathfrak{m}_+$. Notice that this proof relies essentially on the fact that the commutation relations (\ref{commrel}) impose no constraints on the commutator $[\mathfrak{m}_+, \mathfrak{m}_-]$, which is the only nontrivial commutator for the fermions ($[\mathfrak{f}_1, \mathfrak{f}_2]$).

Now, in the case of the real form $su(m_1, m_2|n_1, n_2)$, the fermions $\mathfrak{f}_1, \mathfrak{f}_2$ are complex conjugate to each other: $\mathfrak{f}_2=\,^\ast(\mathfrak{f}_1)$, therefore the definition (\ref{Rferm}) is compatible with this real form. (In our former notations, one can say that $R$ commutes with complex conjugation: $R\,^\ast=\,^\ast R$.) $\blacksquare$

It is clear from the proof that the statement of the lemma is also true if one replaces $su(m_1, m_2|n_1, n_2)$ by $psu(m_1, m_2|n_1, n_2)$ whenever $m_1+m_2=n_1+n_2$ (the commutation relations (\ref{commrelsuper}) are unchanged).

\subsection{Our choice of $R$.}
According to the lemma above, it suffices to pick the action of $R$ on the bosonic subalgebra $su(1,1)\oplus su(2)\oplus u(1)$. First of all, we will require that $R$ acts as a complex structure on $su(2)\oplus u(1)$:
\bea
R\big|_{su(2)\oplus u(1)}=\mathscr{J}(\vec{n})
\eea
As the notation indicates, the complex structure depends on a unit vector $\vec{n}$. As regards the action of $R$ on $su(1,1)$, we will pick the standard one \cite{Hoare}. The generators of $su(1,1)$ in a basis with real structure constants may be taken to be $\sigma_1, \sigma_2, i\sigma_3$. (Here $\sigma_1, \sigma_2, \sigma_3$ are the Pauli matrices.) Then we define
\bea
R(\sigma_1\pm i\sigma_2)=\pm i\, (\sigma_1\pm i \sigma_2), \quad R(i \sigma_3)=0.
\eea
It is clear that $R$ commutes with complex conjugation.

\vspace{0.3cm}
So far we have discussed the choice of $R$-matrix for one copy of the algebra $psu(1,1|2)\oplus u(1)$. The manifold $\mathcal{M}$ has an additional copy of $PSU(1,1|2)\times U(1)$ in its definition (\ref{Mmanifold}) as a quotient space. As discussed in \ref{2param}, this allows, in principle, to set up a further deformation through an $R$-matrix acting on this factor. This possibility was subjected to scrutiny in \cite{Hoare}, the result being that such two-parametric deformation leads to a singular deformation of the $AdS$ part of the background. Such singularity is in fact inevitable in the deformations of higher-dimensional $AdS$ backgrounds, such as in the original case of $AdS_5\times S^5$ \cite{Vicedo, ABF}. The $AdS_3 \times S^3 \times S^1$ background provides a unique opportunity to obtain a smooth deformed space, but this requires precisely setting the second deformation parameter to zero \cite{Hoare}. Embracing this favorable situation, we will henceforth consider only the one-parametric deformation. This will lead, in particular, to the preservation of the left $PSU(1,1|2)\times U(1)$ symmetry.

\subsection{Bosonic part of the deformed background.}
Although the bosonic part of the target space is $SL(2, \mathds{R})\times SU(2)\times U(1)$ -- a group manifold --, the full target space $\mathcal{M}$, incorporating also the fermionic directions, is not. Instead, it is a homogeneos space of a special type -- the Lie algebra $psu(1,1|2)\oplus u(1)\oplus psu(1,1|2)\oplus u(1)$ of its superisometry group is $\mathds{Z}_4$-graded, and the grading-zero component is $su(1,1)\oplus su(2)\oplus u(1)$. The action of an automorphism $\Omega$ of this Lie algebra, which generates the graded structure (i.e. $\Omega^4=1$), is defined to be
\bea\label{Omega}
\Omega (a, b)=(b, (-1)^F a),\quad\quad a\in psu(1,1|2)\oplus u(1),\quad b\in psu(1,1|2)\oplus u(1)\,.
\eea
Here $F$ is the fermion number operator, i.e. $(-1)^F$ is equal to $1$ on a bosonic element of the algebra and $-1$ on a fermionic one. An important property, which ensures that $\Omega$ is a Lie algebra homomorphism, is that $(-1)^F[a_1, a_2]=[(-1)^Fa_1, (-1)^Fa_2]$. The definition (\ref{Omega}) is a fermionic generalization of a $\mathds{Z}_2$-grading, characteristic of symmetric spaces. Indeed, the bosonic part of the target space -- a group manifold $G=SL(2, \mathds{R})\times SU(2)\times U(1)$ -- may be presented as a symmetric space: $G=\frac{G\times G}{G_{diag}}$. The $\mathds{Z}_2$-grading on the Lie algebra $\mathfrak{g}\oplus \mathfrak{g}$ is generated by a permutation: $\Omega(a, b)=(b, a)$, i.e. it is precisely the restriction of the operator $\Omega$ above to the bosonic sector.

Since the target space is not a group manifold, and in fact not even a symmetric space, the general setup of section \ref{general} is not directly applicable. However, there exists a suitable generalization for the case of Green-Schwarz-type models on $\mathds{Z}_4$-graded spaces \cite{Vicedo} \cite{Hoare}. We do not wish to go into these details here and we refer the interested reader to these papers. It is only important that such generalizations do exist, so that it is possible to incorporate the fermions into the construction. For the moment we will concentrate on the bosonic part of the background. Let $g\in PSU(1,1|2)\times U(1)$ be the group element, then the current is
\bear
J:=-g^{-1}dg=\sum\limits_{a=1}^3\,(M)_a\,\lambda^a+\sum\limits_{\mu=0}^3\,(S)_\mu\,\sigma^\mu+\\
+ \sum\;(\textrm{fermionic generators}),
\eear
where $\lambda^1=\sigma^1, \lambda^2=\sigma^2, \lambda^3=i\sigma^3$ are the generators of $su(1,1)$, and, as before, $\{\sigma^\mu\}=i\{\mathds{1}, \vec{\sigma}\}$ are the generators of $u(1)\oplus su(2)$. The bosonic part of the action then coincides with (\ref{action}) and can be written as follows:
\bear
\mathscr{S}=\int\,d^2x\,\left(-(1+\eta^2)\,(M_3)_+ (M_3)_-+(M_1)_+ (M_1)_-+(M_2)_+ (M_2)_-+\right.\\ \nonumber
\left. + \,2\eta\, ((M_2)_-(M_1)_+-(M_2)_+(M_1)_-)+(S_\mu)_+\,(\mathds{1}-\eta\,\mathscr{J})_{\mu\nu}\,(S_\nu)_-
\right)\,.
\eear
Due to the flatness of the current $J$ one has $dM_3\propto M_1\wedge M_2$, so the first two terms in the second line constitute a topological term. It does not affect the classical integrability of the model. To summarize, we have obtained a target space with the metric
\bea
ds^2=-(1+\eta^2) M_3^2+M_1^2+M_2^2+\sum\limits_{\mu=0}^3\,S_\mu S_\mu,
\eea
i.e. a product of a `squashed $AdS$' metric and the unmodified metric on $S^1\times S^3$. There is, however, a nonzero $B$-field on $S^1\times S^3$:
\bea\label{Bfield}
B=\eta\,\mathscr{J}_{\mu\nu} S_{\mu}\wedge S_{\nu}\,.
\eea
Let us analyze the effects of the `squashing' of $AdS$ (This metric was thoroughly analyzed in \cite{GodelAds}. Our parameter $\eta$ is related to the analogous parameter $\mu$ from that paper via $\mu=\sqrt{1+\eta^2}$). They turn out to be far less innocent than what may seem at first glance. In suitable coordinates (see Appendix \ref{Godel}), the squashed metric can be written as follows:
\bea\label{GodelDefMetr}
ds^2=-d\tau^2+d\rho^2+d\psi^2\,(\mathrm{sh}^2\rho-\eta^2 \mathrm{sh}^4\rho)-2\,\sqrt{1+\eta^2}\,\mathrm{sh}^2\rho\,d\psi\,d\tau
\eea
The absence of singularity at $\rho\to 0$ requires that $\psi$ be an angular variable, with period $2\pi$\footnote{Note that the cross-term $2\,\sqrt{1+\eta^2}\,\mathrm{sh}^2\rho\,d\psi\,d\tau$ is also smooth with this identification, as for $\rho\to 0$ one has $\rho^2\,d\psi=xdy-ydx$, $(x, y)$ being the Cartesian coordinates, related to $(\rho, \psi)$ in the standard way.}. The curve with tangent vector ${\dd \over \dd \psi}$, i.e. the one described by fixing $\rho$ and $\tau$ to be constant, has the induced metric $(ds^2)_{\textrm{curve}}=d\psi^2\,(\mathrm{sh}^2\rho-\eta^2 \mathrm{sh}^4\rho)$, which becomes timelike for sufficiently large $\rho$. This implies the existence of closed timelike curves. In fact, the above metric at $\eta=1$ becomes the G\"odel metric \cite{GodelMetr}, known to have this pathological property. One way to avoid closed timelike curves would be to consider $\eta$ purely imaginary, however in that case the $B$-field (\ref{Bfield}) would become imaginary as well.

\section{\Large Discussion}

In the present paper our main goal was to observe a certain relation between the models studied previously by the author \cite{Bykov1}, \cite{Bykov2}, \cite{Bykov3} and the so-called $\eta$-deformations of $\sigma$-models that were introduced in \cite{Klimcik} and applied to the deformations of the $AdS_5 \times S^5$ superstring background (and related backgrounds) in recent years \cite{Vicedo, ABF}.

Let us recapitulate the main similarities and differences between these models. First of all, the model of \cite{Bykov3} may be defined for an arbitrary complex homogeneous target space, and it does not involve any parameters (once the scale of the metric, and the complex structure, have been chosen). In fact, it coincides with the bosonic part of Witten's `topological $\sigma$-model'  \cite{Witten} and inherits the principal feature of the latter, namely the dependence of the Lagrangian on the complex structure of the target space\footnote{One should note however that, since the models we are considering are purely bosonic, there is no sense in which they are `topological'.}. The $\eta$-deformed models, as the name suggests, are in fact continuous deformations of the standard $\sigma$-models with symmetric target spaces. The deformation parameter is customarily called $\eta\in \CC$. Whereas the value of this parameter measures the overall `size' of the deformation, the Lagrangian of the deformed model depends also on a certain operator $R$, which determines the `shape' of the deformation. In order to preserve integrability of the model\footnote{We refer here to the weak form of `integrability', which means that there exists a zero-curvature representation for the equations of motion, or Lax pair.}, this operator has to satisfy an equation called in the literature the `modified classical Yang-Baxter equation'. In the present paper we emphasized that this equation resembles the condition of vanishing of the Nijenhuis tensor, which in turn is a condition for integrability of an almost complex structure. This is the key point which provides a relation between the two classes of models.

For the choice of $R$-matrix mostly studied in the literature, $R$ is however not a complex structure but rather a generalization thereof, satisfying $R^3=-R$. In Section~\ref{limitcase} we showed that the $SU(N)$ principal chiral model deformed by an $R$-matrix of this type, produces, in the limit $\eta\to \pm i$, a $\sigma$-model of the type discussed in~\cite{Bykov3} with target space $SU(N)/S(U(1)^N)$. This limit is, clearly, somewhat degenerate, as it changes the target space of the model, but nevertheless it is well-defined. On the other hand, such a limiting procedure can only produce a target space of the form $G/H$ with \emph{abelian} $H$, whereas the models of \cite{Bykov3} also allow for non-abelian `denominators' $H$ (Such is the case for the manifolds $SU(N)/S(U(n_1)\times U(n_m))$).

In Sections \ref{limitcase}, \ref{complstruct} we pursued further the parallel between the $R$-matrix and the complex structure. In fact, whenever the target space is a compact reductive Lie group of even dimension, there always exists an integrable complex structure on it. Therefore, instead of employing an $R$-matrix satisfying $R^3=-R$, one can require $R$ to be literally the complex structure itself, satisfying $R^2=-1$. With this choice, the corresponding $\eta$-deformed model provides a generalization of the model considered in \cite{Bykov3}. This generalized model can be interpreted in two ways. For imaginary values of $\eta$, this model is well-defined (i.e. has a real action) only for Euclidean signature of the worldsheet -- in this case it is an interpolation between the model of \cite{Bykov3} and the conventional principal chiral model (i.e. the model with a zero $B$-field). For real values of $\eta$, the model is well-defined on a Minkowski worldsheet -- it is a unitary deformation of the principal chiral model. In either case, the deformation amounts to turning on a $B$-field proportional to the (non-closed) K\"ahler form on the target space.

Arguably the simplest nontrivial example is provided by the target space $S^1\times S^3$. We have analyzed this model in Section \ref{S1S3}. Interestingly, $S^1\times S^3$ is also part of the superstring backgrounds $AdS_3\times S^3\times (S^1)^4$ and $AdS_3 \times S^3\times S^3 \times S^1$ \cite{Maldacena}, \cite{Babichenko}. Therefore a natural question arises: do these models also allow complex structure-induced deformations? In a formal sense, the answer is affirmative -- this relies on the results of \cite{Vicedo}, \cite{Hoare}. When it comes to a more physical analysis of the deformation, several difficulties are encountered. First of all, for the deformation to be a string-theoretic deformation, the deformed background would have to satisfy the supergravity equations of motion. In general, this is known not to be the case \cite{FrolovTseytlin}, \cite{WulffTseytlin} (In special cases the deformed backgrounds might still satisfy the supergravity equations, see~\cite{Borsato}). There are, however, even more immediate issues: as a consequence of the non-compactness of the original $AdS$ space subjected to the deformation, the deformed background generally has a naked curvature singularity at a finite proper distance. Potentially the $AdS_3$ case considered in this paper could have been an exception. Since $AdS_3\times S^3$ is a group manifold, it allows a two-parametric family of $\eta$-deformations (the so-called `bi-Yang-Baxter' deformation \cite{Klimcik2}). It was shown in \cite{Hoare} that, when the deformation preserves the left-invariance (or right-invariance) of the background, the curvature singularity is absent. We concentrated on this case in Section \ref{AdSDef}. As we found, however, the deformed background in this case is described by a generalized G\"odel metric with closed timelike curves. The conclusion of the analysis is that the $\eta$-deformations of superstring backgrounds remain dubious, which however should not undermine the value of such deformations for the more conventional asymptotically free $\sigma$-models.

\vspace{0.3cm}
\textbf{Acknowledgements.}
{\footnotesize
I would like to thank S.~Frolov, V.~Pestun, K.~Zarembo, A.~Zotov for discussions. I am grateful to S.~Frolov for comments on the manuscript. I am indebted to Prof.~A.A.Slavnov and to my parents for support and encouragement. My work was supported in part by the grant RFBR 14-01-00695-a.
}

\appendix

\section{The $\eta$-deformed principal chiral model as a gauged  $\sigma$-model}
\subsection{Integrating out the auxiliary gauge field.}\label{integrout}

We start by showing, how the auxiliary gauge field $A_\pm$ may be integrated out of the Lagrangian (\ref{lagrgauged1}). There are at least two ways to do this:

\vspace{0.3cm}
\noindent 1. Straightforward elimination. We rewrite the Lagrangian (\ref{lagrgauged1}) as follows:
\bear\label{lagrgauged}
\widetilde{\mathscr{L}}&=&
\mathrm{Tr}\left((A_+-\mathds{J}_+ \mathcal{S}^{-1})\mathcal{S}(A_--\mathcal{S}^{-1}\mathds{J}_-)\right)-\mathrm{Tr}\left(\mathds{J}_+ \mathcal{S}^{-1} \mathds{J}_-\right)+
\\ \nonumber &&+\mathrm{Tr}\left((J_1)_+\frac{1}{1-\eta_1 (R_1)_{g_1}}\circ(J_1)_-\right)+ \mathrm{Tr}\left((J_2)_+\frac{1}{1-\eta_2 (R_2)_{g_2}}\circ(J_2)_-\right)\,,
\eear
where $(J_i)_\pm=\dd_\pm g_i\,g_i^{-1}$ and
\bear\nonumber
&\mathcal{S}=\frac{1}{1-\eta_1 (R_1)_{g_1}}+\frac{1}{1-\eta_2 (R_2)_{g_2}},\quad\quad R_g:=Ad_g\,R\,Ad_{g^{-1}}\,.&\\ \nonumber
&\mathds{J}_-=\frac{1}{1-\eta_1 (R_1)_{g_1}}\circ (J_1)_-+\frac{1}{1-\eta_2 (R_2)_{g_2}}\circ (J_2)_-,&\\ \nonumber
&\mathds{J}_+=\frac{1}{1+\eta_1 (R_1)_{g_1}}\circ (J_1)_++\frac{1}{1+\eta_2 (R_2)_{g_2}}\circ (J_2)_+&
\eear
Eliminating the first term in (\ref{lagrgauged}), we are left with the following:
\bear\nonumber
\widetilde{\mathscr{L}}&=&
-\mathrm{Tr}\left(\left(\frac{1}{A^T}\circ (J_1)_++\frac{1}{B^T}\circ (J_2)_+\right) \left(\frac{1}{A}+\frac{1}{B}\right)^{-1} \left(\frac{1}{A}\circ (J_1)_-+\frac{1}{B}\circ (J_2)_-\right)\right)+
\\ \nonumber &&+\mathrm{Tr}\left((J_1)_+\frac{1}{A}\circ(J_1)_-\right)+ \mathrm{Tr}\left((J_2)_+\frac{1}{B}\circ(J_2)_-\right)\,,
\eear
where $A=1-\eta_1 (R_1)_{g_1}, B=1-\eta_2 (R_2)_{g_2}$. To simplify the above expression, we will use the identities
\bear\nonumber
&\frac{1}{A}+\frac{1}{B}=\frac{1}{A}(A+B)\frac{1}{B}=\frac{1}{B}(A+B)\frac{1}{A}&\\ \nonumber
&\Rightarrow \frac{1}{A}(\frac{1}{A}+\frac{1}{B})^{-1}\frac{1}{A}=\frac{1}{A}-\frac{1}{A+B},\quad\quad \frac{1}{A}(\frac{1}{A}+\frac{1}{B})^{-1}\frac{1}{B}=\frac{1}{A+B}\,,&
\eear
valid for arbitrary invertible matrices $A, B$.
Using these identities, we obtain 
\bear \nonumber
\widetilde{\mathscr{L}}&=&\mathrm{Tr}\left((J_1)_+ \frac{1}{A+B} (J_1)_-\right)+\mathrm{Tr}\left((J_2)_+ \frac{1}{A+B} (J_2)_-\right)-\\ \nonumber &&-\mathrm{Tr}\left((J_1)_+ \frac{1}{A+B} (J_2)_-\right)-\mathrm{Tr}\left((J_2)_+ \frac{1}{A+B} (J_1)_-\right)=\\ \nonumber
&&=\mathrm{Tr}\left((J_1-J_2)_+ \frac{1}{A+B} (J_1-J_2)_-\right)
\eear
Recalling our definitions of $A$ and $B$, we arrive at the final form of the Lagrangian:
\bea\label{gaugedfinlagr}
\widetilde{\mathscr{L}}={1\over 2}\mathrm{Tr}\left((J_1-J_2)_+ \frac{1}{1-{\eta_1\over 2} R_{g_1}-{\eta_2\over 2} R_{g_2}}\circ (J_1-J_2)_-\right)
\eea
Clearly, this is still gauge-invariant with respect to the transformations $g_1\to g_1 g, g_2\to g_2 g,\;g\in G$. The simplest gauge condition would be to set $g_1=1$ or $g_2=1$.

\vspace{0.3cm}
\noindent 2. Using a Hubbard-Stratonovich transformation. The direct elimination described above is rather cumbersome. There is a simpler way to arrive at the same result. First, we perform a quadratic transformation on the Lagrangian (\ref{lagrgauged1}), introducing new auxiliary fields $\pi_\pm\in \mathfrak{g}, \rho_\pm\in \mathfrak{g}$: 
\bear\label{lagrgauged2}
&&\widetilde{\mathscr{L}}=
-\mathrm{Tr}\left(\pi_+(\mathds{1}-\eta_1 (R_1)_{g_1})\circ\pi_-\right)-\mathrm{Tr}\left(\rho_+(\mathds{1}-\eta_2 (R_2)_{g_2})\circ\rho_-\right)+
\\&&+
\mathrm{Tr}\left(\pi_+(\tilde{J}_1)_-\right)+\mathrm{Tr}\left(\pi_-(\tilde{J}_1)_+\right)\,+\mathrm{Tr}\left(\rho_+(\tilde{J}_2)_-\right)+\mathrm{Tr}\left(\rho_-(\tilde{J}_2)_+\right).
\eear
Here $(\tilde{J}_i)_\pm=\mathscr{D}_\pm g_i\, g_i^{-1}$. The above action is now linear in $A_\pm$. Varying it with respect to $A_\pm$ we get two constraints: $\pi_\pm +\rho_\pm=0$. Therefore we can express the remaining part of the action in terms of $\pi_\pm$:
\bear
&&\widetilde{\mathscr{L}}=-2\, \mathrm{Tr}\left(\pi_+\big(\mathds{1}-{\eta_1\over 2} (R_1)_{g_1}-{\eta_2\over 2} (R_2)_{g_2} \big)\circ\pi_-\right)+\\&&+\mathrm{Tr}\left(\pi_+(J_1- J_2)_-\right)+\mathrm{Tr}\left(\pi_-(J_1- J_2)_+\right)\,.
\eear
We can now eliminate $\pi_\pm$ to obtain the same expression (\ref{gaugedfinlagr}) for the Lagrangian:
\bea
\widetilde{\mathscr{L}}= {1\over 2}\mathrm{Tr}\left((J_1-J_2)_+ \frac{1}{1-{\eta_1\over 2} R_{g_1}-{\eta_2\over 2} R_{g_2}}\circ (J_1-J_2)_-\right)\,.
\eea

\subsection{Derivation of the Lax pair.}\label{laxpair}

We will now derive a Lax pair for the equations of motion following from the Lagrangian (\ref{lagrgauged1}). The variation of (\ref{lagrgauged1}) with respect to $g_1, g_2$ produces the following equations of motion:
\bear\label{eomgroup}
&\mathscr{D}\ast \mathscr{K}_1=0,\quad\quad \mathscr{D}\ast \mathscr{K}_2=0,&\\ \label{noethcurrgroup}
&(\mathscr{K}_1)_\pm={1\over 1\pm\eta_1 R_{g_1}}\circ (D_\pm g_1 g_1^{-1}),\quad\quad (\mathscr{K}_2)_\pm={1\over 1\pm\eta_2 R_{g_2}}\circ (D_\pm g_2 g_2^{-1})&
\eear
Varying the Lagrangian w.r.t. the gauge field $A$, we get an additional constraint
\bea\label{constr}
\mathscr{K}_1+\mathscr{K}_2=0\,.
\eea
Equations (\ref{noethcurrgroup}) may be rewritten in the form (\ref{cartandecomp}):
\bear
\dd_\pm g_1 g_1^{-1}=A_\pm+(1\pm\eta_1 R_{g_1})(\mathscr{K}_1)_\pm,\quad\quad \dd_\pm g_2 g_2^{-1}=A_\pm+(1\pm\eta_2 R_{g_2})(\mathscr{K}_2)_\pm\,.
\eear
Using the flatness of $dg_1 g_1^{-1}$ and $dg_2 g_2^{-1}$, we get two structure equations (\ref{cartaneq}):
\bear
dA-A\wedge A=(1+\eta_1^2) \mathscr{K}_1\wedge \mathscr{K}_1-\mathscr{D} \mathscr{K}_1+\eta_1 \,R_{g_1} (\mathscr{D} \ast \mathscr{K}_1)
\\
dA-A\wedge A=(1+\eta_2^2) \mathscr{K}_2\wedge \mathscr{K}_2-\mathscr{D} \mathscr{K}_2+\eta_2 \,R_{g_2} (\mathscr{D} \ast \mathscr{K}_2)
\eear
Using the equations of motion (\ref{eomgroup}) and the constraint (\ref{constr}), we can express everything in terms of $\mathscr{K}_1:=\mathscr{K}$:
\bear
dA-A\wedge A=(1+\eta_1^2) \mathscr{K}\wedge \mathscr{K}-\mathscr{D} \mathscr{K}
\\
dA-A\wedge A=(1+\eta_2^2) \mathscr{K}\wedge \mathscr{K}+\mathscr{D} \mathscr{K}
\eear
which in turn can be rewritten as
\bear
&dA-A\wedge A=(1+{\eta_1^2+\eta_2^2\over 2}) \mathscr{K}\wedge \mathscr{K}&\\
&\mathscr{D} \mathscr{K}+{\eta_2^2-\eta_1^2\over 2}\,\mathscr{K}\wedge \mathscr{K}=0\,.&
\eear
One can then look for a family of flat connections of the following form:
\bea
\mathscr{A}=A+\alpha \mathscr{K}+\beta \ast \mathscr{K}
\eea
Imposing the flatness condition and using the e.o.m., one arrives at a constraint for $\alpha, \beta$, which can be solved by expressing them in terms of an unconstrained variable $u$. At the end one gets the following expression for the Lax connection:
\bear\nonumber
&\!\!\!\!\!\!\mathscr{A}=A+\left(ab+\sqrt{(1+a^2)(1+b^2)}\, u\right)\,\mathscr{K}_+ dx^++\left(ab+\sqrt{(1+a^2)(1+b^2)}\, {1\over u}\right)\,\mathscr{K}_- dx^-,&\\ \nonumber
&a={\eta_1+\eta_2\over 2}, \quad\quad b={\eta_1-\eta_2\over 2},\quad\quad u\in \CC^\ast.&
\eear

\section{Checking the integrability of complex structures on $S^1\times S^3$}\label{integrcomplstruct}

In this Appendix we check explicitly that the complex structures on $S^1\times S^3$, defined by specifying the holomorphic subspaces (\ref{holspaces}) of the tangent space, are integrable. We will use the representation (\ref{Jn}) for the complex structure (thus assuming that it is anti-self-dual). Then we have
\bea\label{sigmapl}
(\sigma_+)_0=i(\mathds{1}+i \,\vec{n}\cdot\vec{\sigma}),\quad \vec{\sigma}_+=i(\vec{\sigma}+i\, \vec{n}\times \vec{\sigma}-i \,\vec{n}\,\mathds{1})
\eea
Let us now bring the vector $\vec{n}$ to the form $(0, 0, 1)$ by an $SO(3)$ transformation $\Lambda^t$, i.e. $\Lambda^t\circ \vec{n}=(0, 0, 1):=\vec{n}_0$. We can find an element $g\in SU(2)$, such that $g\vec{\sigma} g^\dagger=\Lambda\circ \vec{\sigma}$. Then from (\ref{sigmapl}) we find that
\bear
&&g (\sigma_+)_0 g^\dagger=i(\mathds{1}+i \,\vec{n}_0\cdot\vec{\sigma}),\\ \nonumber
&&g (\Lambda^t\circ\vec{\sigma}_+) g^\dagger = i(\vec{\sigma}+i \Lambda^t\circ (\vec{n}\times (\Lambda\circ \vec{\sigma}))-i\,\vec{n}_0\,\mathds{1})=i(\vec{\sigma}+i \, (\vec{n}_0\times  \vec{\sigma})-i\,\vec{n}_0\,\mathds{1})
\eear
In the last equation we used the fact that, for an arbitrary $\Lambda\in SO(3)$, one has $\Lambda\circ (\vec{a}\times \vec{b})=(\Lambda\circ\vec{a}\times \Lambda\circ\vec{b})$ ($\Lambda$ is an inner automorphism of the Lie algebra $\mathfrak{so}(3)$). It is clear that $(\sigma_+)_0, \Lambda^t\circ\vec{\sigma}_+$ form a basis in the same space $\mg_+$ as the original matrices $(\sigma_+)_0, \vec{\sigma}_+$. Therefore the above formulas imply that, by a $g$-transformation, we have effectively brought the complex structure $\mathscr{J}$ to a canonical form. Indeed, now we can write out explicit expressions for the generators $\tilde{\sigma}_+=\sigma_+|_{\vec{n}=\vec{n}_0}$:
\bear\label{matn0}
& (\tilde{\sigma}_+)_0=\left( \begin{array}{cc}
i-1 & 0   \\
0 & i+1   
 \end{array} \right),\quad (\tilde{\sigma}_+)_1=\left( \begin{array}{cc}
0 & 0   \\
i & 0   
 \end{array} \right), &\\ & (\tilde{\sigma}_+)_2=i (\tilde{\sigma}_+)_1,\quad (\tilde{\sigma}_+)_3= -i\,(\tilde{\sigma}_+)_0 &
\eear
It is now obvious that $[\tilde{\mg}_+, \tilde{\mg}_+]\subset \tilde{\mg}_+$ and hence  $[\mg_+, \mg_+]\subset \mg_+$, so that the complex structure $\mathscr{J}$ is integrable. The case of self-dual $\mathscr{J}$ is analyzed analogously. We have thus arrived at a proof of Lemma 1 from Section \ref{hyperhermit}. Here is a proof of Lemma 2:

\vspace{0.3cm}
\noindent\textbf{Proof of Lemma 2.} It is sufficient to check the vanishing of the $(0, 2)$ components of the metric, i.e. that $\langle\mg_+, \mg_+\rangle=0$. The latter is clear, however, from (\ref{matn0}), as $\tr((\tilde{\sigma}_+)_0^2)=\tr((\tilde{\sigma}_+)_1^2)=\tr((\tilde{\sigma}_+)_0 (\tilde{\sigma}_+)_1)=0$.

\section{The `squashed'-$AdS$/G\"odel metric}\label{Godel}

The group $SU(1,1)$ is defined by the following property: $g\widehat{\eta} g^\dagger=\widehat{\eta}$, where $\widehat{\eta}=\textrm{Diag}(1, -1)$.
An arbitrary element of $SU(1,1)$ can be parametrized by the matrix
\bea
g=\left( \begin{array}{ccc}
z_1 & \bar{z}_2  \\
z_2 & \bar{z}_1  \end{array} \right)
\eea
where $|z_1|^2-|z_2|^2=1$. Let us introduce the `global coordinates' in the following way:
\bea
z_1=\mathrm{ch}\,\rho\,e^{i t},\quad\quad z_2=\mathrm{sh}\,\rho\,e^{i \varphi}
\eea
Then the current is
\bear
&&J:=-g^{-1}dg=-\left( \begin{array}{ccc}
\bar{z}_1 dz_1-\bar{z}_2 dz_2 & \bar{z}_1 d\bar{z}_2-\bar{z}_2 d\bar{z}_1  \\
z_1 dz_2-z_2 dz_1 & z_1 d\bar{z}_1-z_2 d\bar{z}_2  \end{array} \right)=\\ \nonumber &&=\left( \begin{array}{ccc}
i (\mathrm{sh}^2\rho\, d\varphi-\mathrm{ch}^2\rho\, dt) & -e^{-i(t+\varphi)}(d\rho-i\, \mathrm{sh}\rho\, \mathrm{ch}\rho\, (d\varphi-dt))  \\
-e^{i(t+\varphi)}(d\rho+i\, \mathrm{sh}\rho\, \mathrm{ch}\rho\, (d\varphi-dt)) & -i (\mathrm{sh}^2\rho\, d\varphi-\mathrm{ch}^2\rho\, dt)  \end{array} \right)
\eear
The components of the dreibein are:
\bear
&&M_1-iM_2=-e^{-i(t+\varphi)}(d\rho-i\, \mathrm{sh}\rho\, \mathrm{ch}\rho\, (d\varphi-dt)),\\
&&M_3=\mathrm{sh}^2\rho\, d\varphi-\mathrm{ch}^2\rho\, dt
\eear
A simple change of variables $t=\frac{\tau}{\sqrt{1+\eta^2}}$, $\varphi=\frac{\tau}{\sqrt{1+\eta^2}}-\psi$ brings the metric to the form (\ref{GodelDefMetr}):
\bea
ds^2=-d\tau^2+d\rho^2+d\psi^2\,(\mathrm{sh}^2\rho-\eta^2 \mathrm{sh}^4\rho)-2\,\sqrt{1+\eta^2}\,\mathrm{sh}^2\rho\,d\psi\,d\tau\,.
\eea

\vspace{0.7cm}
\begingroup
    \setlength{\bibsep}{4pt}
   \bibliography{S1S3refs}
\bibliographystyle{ieeetr}
\endgroup

\end{document}